\documentclass[preprint,aps,10pt, showpacs,twocolumn]{revtex4}
\begin{document}
\title{The Electronic Structure of CaCuO$_2$ From 
the B3LYP Hybrid Functional}
\author{Xiao-Bing Feng$^{1,2}$ and N. M. Harrison$^{1,3}$
\\
 $^1$Department of Chemistry, Imperial College of Science, Technology\\ 
and Medicine, London, SW7 2AY, UK
\\
 $^2$Department of Physics, Dalian Railway Institute, 
Dalian, 116028, P. R. China
\\
 $^3$CLRC, Daresbury Laboratory, Computational and 
Material Science Department,\\ 
Daresbury, Warrington, WA4 4AD, UK}

\begin{abstract}
    The electronic structure of the infinite layer compound CaCuO$_2$ 
has been calculated with the B3LYP hybrid density functional.
The mixing of the Hartree-Fock exchange in the exchange-correlation energy 
separated the bands crossing Fermi energy to form an antiferromagnetic 
insulating ground state of charge transfer type. The complete elimination 
of the self-interaction through the exact exchange and the optimized 
gradient-corrected correlation energy significantly improved theoretical
results. The theoretical energy gap and magnetic moment are in excellent 
agreement with the experiments. The ratio of intralayer to interlayer 
magnetic coupling constants and lattice parameters are also in good 
accordance with the experiments. Some characteristics of the electronic 
structure of insulating Sr$_2$CuO$_2$Cl$_2$ from angle-resolved 
photoemission experiments are observed in the B3LYP band structure for 
CaCuO$_2$.
\end{abstract}
\maketitle
\par
As a parent compound of high temperature superconductors(HTSC), 
the infinite layer material CaCuO$_2$ \cite{structure} has a 
simple structure and a high transition temperature when doped 
to optimum hole density\cite{tc1,tc2}. Similar to La$_2$CuO$_4$ 
and YBa$_2$Cu$_3$O$_6$, CaCuO$_2$ has a insulating 
three-dimensional antiferromagnetic(AFM) ground state\cite{afm},
with energy gap $\Delta=1.5$eV\cite{gap} and magnetic moment 
$\mu=0.51$ $\mu_B$\cite{afm}. Although the ratio of interlayer 
to intralayer magnetic coupling is one order of magnitude higher 
than that in La$_2$CuO$_4$ and YBa$_2$Cu$_3$O$_6$, the 
ratio is still very low. The nuclear magnetic resonance 
experiment on Cu in Ca$_{0.85}$Sr$_{0.15}$CuO$_4$ shows that the 
interlayer coupling is about two orders of magnitude less than 
the intralayer one\cite{coupling}.  
\par
The electronic structure of CaCuO$_2$ has been investigated by 
several theoretical methods, such as LAPW\cite{lapw}, LMTO-ASA\cite{lmto-asa}, 
FLMTO\cite{flmto}. These LSDA-based methods failed to give the 
correct ground state, as in the case of other HTSCs \cite{rmp}.  
Later, it was found that the failure was due to the self-interaction 
inherent in the LSDA density functional, which tends to delocalize 
the electrons. The self-interaction correction(SIC) method gave 
correct AFM insulating ground state with $\mu=0.58$ $\mu_B$ and 
$\Delta=0.84$ eV\cite{sic-cacuo2}. The LSDA+$U$ method also generated 
correct ground state, with $\Delta=2.1$ eV and $\mu=0.66$ $\mu_B$ for 
$U=7.5$ eV\cite{lsda+u1} and $\Delta=1.96$ eV and $\mu=0.71$ $\mu_B$ 
for $U=5$ eV\cite{lsda+u2}. Although the two approaches recovered 
the correct ground state, their quantitative comparisons with 
the experiments were not satisfactory.
\par
In this paper we study the electronic structures of CaCuO$_2$ 
with the so-called B3LYP hybrid density functional. The hybrid 
functional was very successful in the thermochemistry of atoms 
and molecules\cite{becke}. Later, the hybrid functional was 
applied to some periodic systems\cite{b3lypapp1,b3lypapp3}. 
The argument for mixing the  Hartree-Fock(HF) exchange 
in the exchange-correlation energy $E_{XC}$ is based on the 
adiabatic connection formula\cite{becke},
\begin{equation}
\label{eq1}
E_{XC}=\int_0^1 U^\lambda_{XC} d\lambda,
\end{equation}
where $\lambda$ is the electron-electron interaction parameter, 
with $\lambda=1$ corresponding to the real interaction and 
$\lambda=0$ a system of noninteracting electrons. $U^\lambda_{XC}$ 
is the potential energy of the exchange-correlation at any 
$\lambda$. In the case of $\lambda=0$ $U^0_{XC}$ is the HF 
exchange energy of the corresponding noninteracting system. 
The weights for the gradient-corrected correlation energy, 
local exchange energy and the exact HF exchange terms were 
determined by a linear least-square fitting of the thermochemical 
properties of some atoms and molecules. The atom with highest 
atomic number used in the fitting is Cl. No atoms with d or 
higher shells were used. $20\%$ mixing of the exact HF exchange 
energy in the $E_{XC}$ gives theoretical results in good 
agreement with experiments. In the B3LYP scheme the 
Perdew-Wang\cite{pw91} gradient-corrected correlation energy 
is replaced by Lee-Yang-Parr correlation energy\cite{lyp}. 
\par
The final exchange-correlation energy functional reads
\begin{eqnarray}
\label{hybrid}
E_{XC}=&E^{LSDA}_X (1-a_0) + a_0 E^{exact}_X 
              +a_X \Delta E^{B88}_X\nonumber\\ 
       &+E_C^{LSDA} + a_C\Delta E_C^{LYP},
\end{eqnarray}
in which the local spin density functional of Vosko, Wilk and 
Nusair\cite{vwn} is used for $E^{LSDA}_X$ and $E^{LSDA}_C$. 
$E_X^{exact}$ is the exact nonlocal HF exchange energy.  
$\Delta E^{B88}$ and $\Delta E_C^{LYP}$ are the Becke's\cite{b88} 
and Lee-Yang-Parr's gradient corrections for the local exchange 
and correlation energies, respectively. The optimum values for 
the parameters $a_0$, $a_X$ and $a_C$ are 0.20, 0.72, and 0.81, 
respectively\cite{becke}. 
\par
In this paper the calculations are carried out with CRYSTAL 
package\cite{crystal98}. The basis vectors for expanding the 
Kohn-Sham orbitals are Bloch functions composed of localized 
contracted basis sets\cite{bs}. All-electron basis sets for Ca, 
Cu and O ions are of the form of 86-511G, 86-411(41d)G and 
8-411G, respectively. 
\par
In the calculations 75 points in the irreducible part of the 
first Brillouin zone(FBZ) were used.  We adopt the  7, 7, 7, 
7 and 14 as the integral tolerances to obtain high precision 
in monoelectronic and bielectronic integrals. Also a strict 
criteria for convergence, i.e. when the difference between two 
consecutive root-mean-squared values of the density matrix 
elements is less than $10^{-9}$ the convergence is assumed, 
is used to make the magnetic moments on Cu sites converge. 
Different magnetic states, i.e. metallic(without spin polarization), 
ferromagnetic, two-dimensional(2D) and three-dimensional(3D) 
antiferromagnetic structures, were checked to find the ground 
state magnetic configuration. For comparisons, the unrestricted 
Hartree-Fock(UHF) and general gradient-corrected LSDA(LSDA+GGA) 
calculations were also carried out with the same basis sets. 
We also performed structural optimization for the 3D AFM supercell.
The optimization is implemented with Domin algorithm and 
the P4/mmm symmetry is preserved. 
\par
Our results show that the 3D AFM state has the lowest energy, 
i.e. the B3LYP hybrid functional could predict correct ground state 
magnetic configuration. The energy difference per unit cell between 
the ferromagnetic and the 2D AFM(antiferromagnetic order in the 
CuO$_2$ plane and ferromagnetic order in the Z direction) ground 
state is 0.285 eV. The 2D and 3D AFM states has nearly the same 
energies, with the energy of 3D AFM state is only 0.00164 eV lower than 
that of 2D AFM state. The energies of different magnetic configurations 
could be described by the Ising model, 
\begin{equation}
E=\sum_{<ij>} J_{ij} S_i S_j.
\end{equation}
The intralayer coupling constant $J_{\|}$ and interlayer coupling 
constant $J_{\bot}$ could be obtained from the energy differences 
of the three different magnetic states. One can easily get 
$J_{\|}=[E_{FM}-E_{2D AFM}]/4$ and $J_{\bot}=[E_{2D AFM}-E_{3D AFM}]/2$,
in which $E_{FM}$, $E_{2D AFM}$ and $E_{3D AFM}$ are the energies 
of FM, 2D AFM and 3D AFM states, respectively. The theoretical results 
are $J_{\|}=825.5$K and $J_{\bot}=9.49$K. The ratio $J_{\bot}/J_{\|}$ 
is 1.1\%, which is in good agreement with the experimental result that 
the interlayer magnetic coupling is about 2 orders of magnitude weaker 
than the intralayer coupling\cite{coupling}.
\par
The optimized lattice parameters for the supercell are a=5.563{\AA} 
and c=6.536{\AA}. The results are about 2\% larger than the experimental 
values\cite{structure}. The precision is comparable to LDA 
computations for conventional materials. In addition, Table~\ref{table1} 
shows that the B3LYP theoretical magnetic moments and energy gap 
obtained with experimental and theoretical lattice parameters are 
in excellent agreement with the experiments\cite{afm,gap}.  As 
mentioned above the optimum parameters were obtained with atoms and 
molecules without d-electrons, so the excellent agreement between 
the theory and experiments was unexpected because of the strong 
localized character of d orbitals. But from the study of 
pseudopotentials one knows that the 2p orbitals in the first-row 
atoms also pose problems, these orbitals have also localized character 
because of the lack of orthogonal repulsions. So, if there are enough 
variational freedoms in the B3LYP hybrid functional to correctly 
describe atoms with 2p shell then it could be applied to  systems 
with 3d electrons. 
\begin{table}
\caption{\label{table1} The energy gaps $\Delta$(in eV) and magnetic 
moments $\mu$(in $\mu_B$) from different theoretical schemes are 
compared with experiments for CaCuO$_2$ and La$_2$CuO$_4$. The results 
with superscripts a and b are obtained using experimental and optimized 
lattice parameters, respectively.}
\begin{ruledtabular}
\begin{tabular}{l|c|c|c|c|c}
              &UHF
              &SIC-LSD
              &LSDA+$U$
              &B3LYP
              &Expt \\
\hline
CaCuO$_2$ ($\Delta$)    &14.9   &0.84$^{[11]}$
              &2.1$^{[12]}$  &1.54$^a$    
              &1.5$^{[5]}$      \\
              &          &         &1.96$^{[13]}$
              &1.50$^b$   &            \\
\hline
CaCuO$_2$ ($\mu$)      &0.89  &0.58$^{[11]}$ 
              &0.66$^{[12]}$ &0.51$^a$ 
              &0.51$^{[4]}$\\
              &   & &0.71$^{[13]}$
              &0.51$^b$  &\\
\hline
La$_2$CuO$_4$ ($\Delta$)     &17$^{[27]}$  
              &2.1$^{[28]}$
              &1.65$^{[29]}$  &2.0$^{[16]}$ 
              &2.0$^{[30]}$\\
              & & 1.04$^{[26]}$ & & & \\
\end{tabular}
\end{ruledtabular}
\end{table} 
\par
The large energy gap of the UHF results from the lack
of correlation, i.e. the interaction is not screened. As can
be seen in Table~\ref{table1} that the UHF gaps for both
CaCuO$_2$ and La$_2$CuO$_4$ are about one order of magnitude
larger than the experimental results. It is also much larger
than the usual cases, which means that the correlation effects are
more important in these materials.
\par
The projected densities of states(DOSs) are shown in Fig.~1 
and the band structures along the high symmetry lines are 
shown in Fig.~2(a). These results are obtained with experimental 
structural parameters. The results show that CaCuO$_2$ is a 
charge transfer insulator. This is similar to the case of 
transition metal monoxides\cite{b3lypapp1}. As shown in Fig.~1 
the main components at the top of the valence bands and the 
bottom of the conduction bands are Cu 3d$_{x^2-y^2}$ and 
O 2p$_{x(y)}$. Though the spectral weight at the bottom of 
the conduction bands is mostly of Cu 3d$_{x^2-y^2}$ nature, 
there is a significant ingredient of O 2p$_{x(y)}$ at the 
top of the valence bands. Because of the lack of apical 
oxygens in CaCuO$_2$ the bonding along the Z direction is weak, 
which is different from La$_2$CuO$_4$, where significant 
Cu 3d$_{z^2}$/O 2p$_z$ components are existent at the top 
the valence bands\cite{b3lypapp3}. The result suggests that 
the Cu 3d$_{z^2}$/O 2p$_z$ components in La$_2$CuO$_4$ may 
not be essential to the hight temperature superconductivity 
of the system. 
\par
One may also notice that some features in Fig.~2(a) are in 
good agreement with the angle-resolved photoemission(ARPES) 
experiments on the insulating Sr$_2$CuO$_2$Cl$_2$\cite{ARPES}.
Sr$_2$CuO$_2$Cl$_2$ has a structure similar to La$_2$CuO$_4$, 
with the apical oxygens in La$_2$CuO$_4$ replaced by 
Cl\cite{SCOC_structure}. The X and M points in Fig.~2(a)
correspond to the ($\pi/2$, $\pi/2$) and ($\pi$, 0) points 
in the FBZ of the CuO$_2$ plane, respectively. The ($\pi$, $\pi$) 
point in the ARPES experiments is equivalent to $\Gamma$ point 
in Fig.~2(a). The ARPES experiments show that the valence band 
top is at ($\pi/2$, $\pi/2$) and there is a nearly isotropic 
dispersion around this point. The dispersion from ($\pi$, 0) 
to (0, 0) is flat and the dispersion from ($\pi$, 0) to 
(0, $\pi$) is similar to the one from (0, 0) to ($\pi$, $\pi$). 
All these characteristics are consistent with our results,
as can be seen in Fig.~2(a).  Due to the folding-up of the 
FBZ in the case of the AFM supercell, there are two nearly 
degenerate bands around X points. Although the dispersion along 
the $\Gamma$-X-M direction in Fig.~2(a) is stronger than 
experimentally observed, a better agreement with experiments 
may be expected by applying the method to Sr$_2$CuO$_2$Cl$_2$. 
\par
To see the effect of the mixing of the exact exchange the LSDA+GGA 
band structure is shown in Fig.~2(b) for comparison. The same 
supercell and  same functionals for GGA corrections were used 
in the LSDA+GGA and the B3LYP calculations.  From Fig.~2(a) and Fig.~2(b) 
one can see that the mixing of the exact exchange results in no 
significant change of the dispersions and relative positions 
of the bands far from Fermi energy $E_F$. But the important 
effect of mixing exact exchange is that two bands, which 
cross $E_F$, are separated from the other two bands. The 
separated two bands have no overlap with the other conduction 
bands, which is different from the LSDA+$U$ results\cite{lsda+u1,
lsda+u2}. In the LSDA+$U$ scheme there are also two separated 
bands composed of Cu 3d$_{x^2-y^2}$ and O 2p$_{x(y)}$ antibonding 
states, but the two bands are overlapped with other conduction 
bands\cite{lsda+u2}. 
\par
Except the difference mentioned above between the B3LYP and 
the LSDA+$U$ schemes, some essential characters are shared by 
the two approaches. The reduced energy gap results from upward 
shift of part of Cu 3d$_{x^2-y^2}$ and O 2p$_{x(y)}$ spectral weights, 
and this part of spectral weights have similar width in both 
schemes. Also the most important projected DOSs for Cu 
3d$_{x^2-y^2}$ and O 2p$_{x(y)}$ have the same 
characters\cite{lsda+u1}. Although the LSDA+$U$ captures the 
essential physics in strongly correlated systems its quantitative 
results are not as good as the B3LYP. The reason may be due to the 
simple mean-field treatment of the Hubbard $U$ term in the LSDA+$U$. 
Recently, Mazin \textsl{et al.} have shown that the LSDA+$U$  
failed for moderately-correlated metal, such as FeAl\cite{failure}.
\par
In the LSDA+$U$ scheme it is easy to see the reason why the 
bands crossing $E_F$ were separated to form a gap. The additional 
orbital dependent potential splits a band into Hubbard subbands, 
with separation equal to approximately the screened on-site 
Coulomb repulsion $U$. The SIC approach restores the localized 
character of d orbitals, thus introducing strong on-site Coulomb 
repulsion. In the studies of the Mott insulators it was found that 
the simple UHF can give qualitatively correct results, whereas 
more advanced DFT methods failed\cite{uhf}. The reason is that 
in the UHF there is no self-interaction. 
\par
In our UHF calculation for CaCuO$_2$ the energy gap is about one 
order of magnitude larger than the experimental result. It's 
expectable that small amount of mixing of the exact exchange 
would reduce the unphysical large gap to the experimental value 
while maintaining the correct ground state magnetic configuration. 
But in the B3LYP approach it is not very clear why only $20\%$ 
mixing of the exact exchange into the GGA corrected LSDA can greatly 
change the LSDA+GGA results.  To see the point one can rewrite 
Eq.~(\ref{hybrid}) as
\begin{eqnarray}
\label{eq3}
E_{XC}=&& E^{exact}_X +(1-a_0)(E_X^{LSDA}-E_X^{exact}) \nonumber\\
       && +a_X\Delta E_X^{B88} +E_C^{LSDA} +a_C \Delta_C^{LYP}.
\end{eqnarray}
Instead of only $20\%$ exact exchange one can view the B3LYP as 
incorporating $100\%$ exact exchange, and treat the second and 
the third terms as additional contributions to the correlation 
energy. So, with the exact exchange, where the self-interaction is 
completely excluded, and additional variation freedoms for 
correlation energy one should expect that the method would give 
better results than the LSDA+GGA and the UHF approaches, in the 
latter case  the correlation energy is totally discarded. 
\par
One may view the B3LYP hybrid functional as a much improved 
energy functional towards the exact one. For a specific class 
of materials better results can be obtained by reoptimizing
the parameters appearing in the correlation energy and by 
adding additional functionals. It is also interesting to test 
if the B3LYP functional could be successfully applied to
moderately-correlated and doped strongly correlated systems. 
Muscat \textsl{et al.} have applied the B3LYP to materials 
of different types of bonding\cite{b3lypapp1}. The theoretical 
energy gaps are in good agreement with experiments, and the 
band structure for Si is also in good accordance with experiments 
and other theoretical methods, such as quantum Monte Carlo. 
For the Mott insulator NiO the B3LYP also generated 
much better energy gap than the LSDA+$U$\cite{lsda+u-nio} and 
the SIC-LSD\cite{sic-la2142,lsda+u-nio}. So, one can expect more 
successful applications of the method to other materials. 
\par
To conclude, the B3LYP scheme has been applied to CaCuO$_2$. 
The mixing of the exact exchange eliminated the self-interaction 
and caused the band separation about the Fermi energy compared 
with the LSDA+GGA method. The magnetic moment and energy gap 
are in excellent agreement with experiments. Some of the 
structural and magnetic properties are also in good agreement 
with experiments. The ARPES characteristics on 
Sr$_2$CuO$_2$Cl$_2$ are observed in our calculations on CaCuO$_2$.

\newpage
\vskip 2cm
\centerline{\Large{\textbf{Figure Captions}}}

\vskip 1cm

\textbf{Fig. 1}
The projected densities of states of Cu 3d, O 2p and Ca 
3sp partial waves in a $\sqrt{2}\times\sqrt{2}\times 2$ 
supercell of CaCuO$_2$. All the DOSs are for spin-up 
electrons and the magnetic moment of Cu is -0.51$\mu_B$.

\vskip 1cm

\textbf{Fig. 2}
(a). The B3LYP energy bands of a $\sqrt{2}\times\sqrt{2}
\times 2$ supercell of CaCuO$_2$. (b). The LSDA+GGA 
band structure of the same supercell. The coordinates 
of the high symmetry points are $\Gamma=(0, 0, 0)/2$, 
$M=(1, 1, 0)/2$, $Z=(0, 0, 1)/2$, $R=(0, 1, 1)/2$ 
and $A=(1, 1, 1)/2$. The bands are for spin-up electrons.


\begin{thebibliography}{99}
%
\bibitem{structure} T. Siegrist \textsl{et al.}, Nature 
\textbf{334}, 231(1988).
%
\bibitem{tc1} J. H. Schon \textsl{et al.}, Nature 
\textbf{414}, 434 (2001). 
%
\bibitem{tc2}M. Azuma \textsl{et al.}, Nature 
\textbf{356}, 775 (1992).
%
\bibitem{afm}D. Vaknin \textsl{et al.}, Phys. Rev. 
B \textbf{39}, 9122 (1989).
%
\bibitem{gap} Y. Tokura, S. Koshihara, and T. Arima, 
Phys. Rev. B \textbf{41}, 11657 (1990).
%
\bibitem{coupling}A. Lombardi \textsl{et al.}, 
Phys. Rev. B \textbf{54}, 93 (1996).
%
\bibitem{lapw}L. F. Mattheiss and D. R. Hamman, 
Phys. Rev. B \textbf{40}, 2217 (1989); 
D. Singh \textsl{et al.}, Physica B \textbf{163}, 470 (1990).
%
\bibitem{lmto-asa} M. A. Korotin and V. I. Anisimov, 
Mater. Lett. \textbf{10}, 28 (1990).
%
\bibitem{flmto} D. L. Novikov, V. A. Gubanov, 
and A. J. Freeman, Physica C \textbf{210}, 301 (1993).
%
\bibitem{rmp} W. E. Pickett, Rev. Mod. Phys. 
\textbf{61}, 433 (1989).
%
\bibitem{sic-cacuo2}D. Singh, W. E. Pickett, and 
H. Krakauer, Physica C \textbf{162-164}, 1431 (1989).
%
\bibitem{lsda+u1}Vladimir I. Anisimov, Jan Zaanen, 
and Ole K. Andersen, Phys. Rev. B \textbf{44}, 943 (1991). 
%
\bibitem{lsda+u2}Hua Wu \textsl{et al.}, 
J. Phys.: Condens. Matter \textbf{11}, 4637 (1999).
%
\bibitem{becke}Axel D. Becke, 
J. Chem. Phys. \textbf{98}, 5648 (1993).
%
\bibitem{b3lypapp1}J. Muscat, A. Wander, N. M. Harrison, 
Chem. Phys. Lett. \textbf{342}, 397 (2001).
%
%
\bibitem{b3lypapp3} J. K. Perry, J. Tahir-Kheli, and 
W. A. Goddard, Phys. Rev. B \textbf{63}, 144510(2001).
%
\bibitem{pw91} J. P. Perdew, in \textsl{Electronic 
Structure of Solids}, edited by P. Ziesche and H. Eschrig 
(Academic Press, Verlag, Berlin, 1991).
%
\bibitem{lyp}Chengteh Lee, Weitao Yang, and Robert G. Parr, 
Phys. Rev. B \textbf{37}, 785 (1988).
%
\bibitem{vwn}S. H. Vosko, L. Wilk, and M. Nusair, 
Canadian J. Phys. \textbf{58}, 1200 (1980).
%
\bibitem{b88}A. D. Becke, Phys. Rev. A \textbf{38}, 3098 (1988).
%
\bibitem{crystal98} V. R. Saunders \textsl{et al.}, 
CRYSTAL98 User's Manual (University of Torino, Torino, 1998).
%
\bibitem{bs}http://www.chimifm.unito.it/teorica/crystal/Basis\_Sets/ 
mendel.html
%
\bibitem{failure} I. I. Mazin \textsl{et al.}, 
arXiv:Cond-mat/0206548.
%
\bibitem{uhf} N. M. Harrison \textsl{et al.}, in \textsl{
The Metal-Non Metal Transition in Macroscopic and Microscopic 
Systems} \textbf{356} (1735) p.75 P. P. Edwards \textsl{et al.} 
(Eds), Phil. Trans. A., (1997)
%
\bibitem{lsda+u-nio} S. L. Dudarev \textsl{et al.}, 
Phys. Stat. Sol. (a) \textbf{166}, 429 (1998).
%
\bibitem{sic-la2142} A. Svane, Phys. Rev. Lett. 
\textbf{68}, 1900 (1992).
%
\bibitem{uhf-la214} Y. -S. Su \textsl{et al.}, 
Phys. Rev. B \textbf{59}, 10521 (1999).
%
\bibitem{sic-la2141}W. M. Temmerman, Z. Szotek, 
and H. Winter, Phys. Rev. B \textbf{47}, 11533 (1993).
%
\bibitem{lsda+u-la214} M. T. Czyzyk and G. A. Sawatsky, 
Phys. Rev. B \textbf{49}, 14211 (1994).
%
\bibitem{la-gap}J. M. Ginder \textsl{et al.}, 
Phys. Rev. B \textbf{37}, 7506 (1988).
%
\bibitem{ARPES}B. O. Wells \textsl{et al.}, Phys. Rev. 
Lett. \textbf{74}, 964(1995); C. Kim \textsl{et al.}, 
\textsl{ibid.} \textbf{80}, 4245(1998).
%
\bibitem{SCOC_structure}L. L. Miller \textsl{et al.},  
Phys. Rev. B \textbf{41}, 1921(1990).
%
\end{thebibliography}
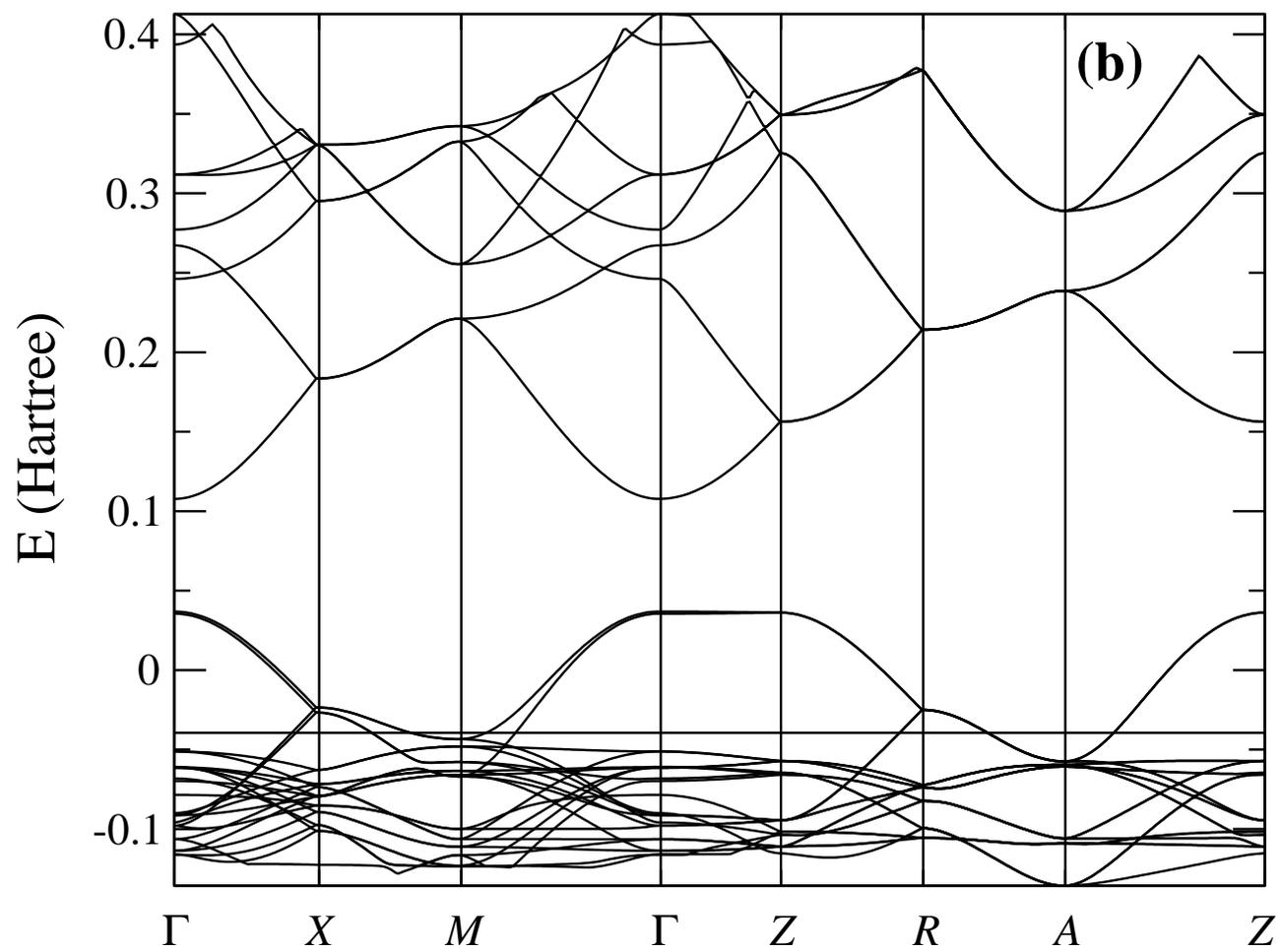
\end{document}